\documentstyle[12pt,epsf]{article}
\newcommand{\be}{\begin{equation}}
\newcommand{\ee}{\end{equation}}
\newcommand{\bea}{\begin{eqnarray}}
\newcommand{\eea}{\end{eqnarray}}
\newcommand{\un}{\underline}

\newcommand{\half}{\raisebox{.15ex}{\scriptsize$\frac{1}{2}$}}

\newcommand{\fourth}{\raisebox{.15ex}{\scriptsize$\frac{1}{4}$}}

\newcommand{\ethpi}{\raisebox{.15ex}{\scriptsize$\frac{1}{8\pi}$}}

\newcommand{\hpi}{\raisebox{.15ex}{\scriptsize$\frac{1}{2\pi}$}}
\newcommand{\fstpi}{\raisebox{.15ex}{\scriptsize$\frac{1}{\pi}$}}
\newcommand{\dmu}{\partial_{\mu}}
\newcommand{\dnu}{\partial_{\nu}}
\newcommand{\re}{\mbox{I$\!$R}}
\newcommand{\co}{\mbox{$l\!\!\!$C}}
\newcommand{\na}{\mbox{I$\!$N}}
\newcommand{\zahlen}{{\rm Z \!\! Z}}

\newcommand{\cL}{\mbox{$\cal L$}}
\newcommand{\cO}{\mbox{$\cal O$}}
\newcommand{\cH}{\mbox{$\cal H$}}

\newcommand{\refeq}[1]{\mbox{eq.~(\ref{eq:#1})}}

\newcommand{\nlsig}{\mbox{O$(3)\,\sigma$-model}}
\newcommand{\sutw}{\mbox{SU$(2)$}}
\newcommand{\sun}{\mbox{SU$(N)$}}
\newcommand{\soth}{\mbox{SO$(3)$}}
\newcommand{\tore}{\mbox{${\rm T}^1\times\re$}}
\newcommand{\torekl}{\mbox{\scriptsize${\rm T}^1\times\re$}}
\newcommand{\toto}{\mbox{${\rm T}^1\times {\rm T}^1$}}

\newcommand{\thrto}{\mbox{${\rm T}^3$}}
\newcommand{\threetore}{\mbox{${\rm T}^3\times\re$}}
\newcommand{\threetoto}{\mbox{${\rm T}^3\times {\rm T}^1$}}

\newcommand{\intT}{\mbox{$[-\half T, \half T]$}}
\newcommand{\vn}{\mbox{$\vec{n}$}}
\newcommand{\vnx}{\mbox{$\vec{n}(x)$}}
\newcommand{\intx}{\mbox{$\int\!\!d^2x\,\,$}}
\newcommand{\intxD}[1]{\mbox{$\int_{#1}\!\!d^2x\,\,$}}
\newcommand{\bu}{\mbox{$\bar{u}$}}
\newcommand{\dz}{\mbox{$\partial_z$}}
\newcommand{\dbz}{\mbox{$\partial_{\bar{z}}$}}

\begin{document}

\hfill INLO-PUB-5/94\\
\begin{center}
{\LARGE{\bf{Tunneling through sphalerons:}}}\\
{\Large{\bf{\underline{the \nlsig\ on a cylinder}}}}\\
\vspace*{1cm}
Jeroen Snippe\footnote{E-mail: snippe@rulgm0.LeidenUniv.nl}\\
\vspace*{1cm}
Instituut-Lorentz for Theoretical Physics,\\
University of Leiden, PO Box 9506,\\
NL-2300 RA Leiden, The Netherlands.\\
\end{center}

\vspace*{5mm}\noindent
\un{Abstract:}

We construct all instantons for the \nlsig\ on a cylinder,
known not to exist on a finite time interval.
We show that the widest instantons go through sphalerons. A
re-interpretation of moduli-space transforms the scale parameter
$\rho$ to a boundary condition in time. This may give a handle on
the $\rho\rightarrow0$ divergent instanton gas.

\section{Introduction}

Due to asymptotic freedom, the large distance behavior of
\sun~gauge theories must be treated non-perturbatively.
A convenient method is to put the model in a finite spatial box
of length $L$ and calculate the low-lying energy eigenstates as
function of $L$. Obviously the wave functionals of
such states are concentrated at small $L$
in the vacua of {\em classical}
configuration space. At larger $L$ they can spread out over low energy
barriers between these vacua. This spreading causes the breakdown
of conventional perturbation theory.

This picture has been used~\cite{LuKoBa,BaDa} for reduction of the
field theory to a finite dimensional system. The remaining degrees
of freedom are expected to correspond to the set of vacua (called
the vacuum valley) and suitable paths over the barriers between
vacua. One of the requirements on such paths is that they cross
a barrier at its lowest point. This point, a sphaleron~\cite{Ta,KlMa},
is defined through a mini-max procedure. One first finds the maximal
energy on a path connecting two vacua. Then one minimizes this maximum
over the space of all such paths.
In local terms a sphaleron is
a saddle point of the energy functional with exactly one
unstable mode.

Well-known paths between two vacua are instantons~\cite{BePo,BePoScTy},
interpreting Euclidean time as a path parameter. In the
WKB-approximation the tunneling amplitude is dominated by paths near
instantons.
Therefore it seems natural to assume that amongst the instanton paths
there is one which crosses a barrier at its lowest point, i.e. it
goes through a sphaleron. Let us assume that the set of instantons
(called the instanton moduli-space) has a scale parameter $\rho$.
Since all instantons have the same Euclidean action, a dimensional
argument easily shows that the maximal energy along the tunneling
path goes as $1/\rho$. Hence it is likely that instantons with
maximal width, as set by $L$, go through a sphaleron.

The latter assumption was made in~\cite{BaDa} for the following
reasons.
First, in a direct approach it is often easier to find
instantons than sphalerons. Then, if the assumption holds,
one can find all sphalerons via instantons. Second,
for small $L$ (high barriers) the spreading of a
low energy wave function is determined by tunneling. For large
$L$ (low barriers) it is determined by classically allowed motion
through a sphaleron. If the
assumption does not hold, there is some range of $L$~values in
which the most important paths between vacua change dramatically.

The conjecture that a sphaleron lies on an instanton path is
{\em in general} not true. For example, consider the two-dimensional
potential $V(q_1,q_2)=(q_1^2-1)^2\mbox{$((q_2^2-1)^2(2+q_2^2)+1)$}
+(1+b^2/q_2^2)^{-1}$ for small values of the parameter $b$.
It has vacua at $(\pm 1, 0)$  with zero energy
and (to leading order in $b$) sphalerons
at $(0,\pm 1)$ with energy 2. However, for small enough $b$
the instanton does not go through a sphaleron. Instead,
it goes straight through the saddle point at $(0,0)$, which has
one unstable mode and energy 3. It is even possible that an
instanton goes through several saddle points with one unstable
mode (we think of a one-dimensional example), in which only the
one with {\em highest} energy would be a candidate sphaleron.

The above clearly shows that in field theory the conjecture
needs to be checked. For \sutw~gauge theory on a space-time
$\thrto\times\intT$ we~\cite{PeGoSnBa}
have developed a numerical procedure to find the widest
instantons (for $T\rightarrow\infty$). Subsequent numerical
investigation~\cite{GaBa} has shown that the widest instanton
goes through a saddle point with one unstable mode, like for
\sutw\ on a space-time $S^3\times\re$~\cite{BaDa}. It is
likely these saddle points are sphalerons. For the two-dimensional
\nlsig, which is studied in the present paper, this can be
proven rigorously.

It is well-known that this $\sigma$-model shares with the
four-dimensional gauge theories features like renormalizability and
asymptotic freedom. Another
similarity is still more important to us now. Both models,
when put in a spatial cube with periodic boundary conditions,
have vacuum valleys of nonzero dimension. This
can increase the number of
instantons, as they
must have endpoints in the vacuum valley. In this context
we also mention the
absence of instantons on compact
space-times \threetoto\ and \toto,
respectively~\cite{BrBa,RiRo,MiSp}. This does not rule out
the existence of instantons on an
infinite time interval. It merely suggests the vacuum valley
cannot be reached in a finite amount of time.

The outline of this paper is as follows: in
section~\ref{sec:genform} we will set up a convenient formalism,
which will be used in
section~\ref{sec:stat} to derive all static solutions of the
Euler-Lagrange equations. In particular we
find the vacuum valley and the
sphaleron solutions. Section~\ref{sec:inst} is devoted
to constructing all instanton solutions and interpreting their
moduli-space. A scale parameter will emerge that relates the
instanton field at $t=+\infty$ to that at $t=-\infty$.
We conclude in section~\ref{sec:concl} by
putting our results in
perspective with respect to the four-dimensional \sutw\ case.

\section{The \nlsig\ in general coordinates}
\label{sec:genform}

The action for the \nlsig\ on a cylinder reads
\be
S[\vn]=\half\intx |\dmu\vnx|^2,\ \ \ \vnx\in\re^3,\ \ \
|\vnx|^2=1,
\label{eq:nact}
\ee
where the integration runs over space-time
$\{(x_2,x_1)\in\tore\}$. We use
overall scale invariance to fix the length of the spatial 1-torus
${\rm T}^1$ (the circle) to be $2\pi$. So $\vn(x+2\pi\hat{e}_2)=\vn(x)$.
The metric on space-time is Euclidean, and we use the summation
convention over repeated indices throughout.

By definition, $\vnx\in S^2$. If we use coordinates $v^i$ ($i=1,2$)
and a metric $g_{ij}$ on $S^2$, \refeq{nact} can be
rewritten as
\be
S[v]=\half\intx g_{ij}(v(x))\dmu v^i(x)\dmu v^j(x).
\label{eq:vact}
\ee
In this paper we will use both
spherical coordinates $(\vartheta,\varphi)$ and
stereographic projection $(u_1,u_2)$ which will be paired as
$u\equiv u_1+i u_2$, with complex conjugate $\bar{u}=u_1-i u_2$
(c.f.~\cite{BePo,RiRo}):
\be
\vn=\left(\!\!\begin{array}{c}\sin\vartheta\cos\varphi \\
\sin\vartheta\sin\varphi \\ \cos\vartheta\end{array}\!\!\right)=
\frac{1}{1+|u|^2}\left(\!\!\begin{array}{c}u+\bu \\ (u-\bu)/i
\\ |u|^2-1\end{array}\!\!\right)
\label{eq:npar}
\ee
(hence $u=\cot\half\vartheta\,e^{i\varphi}$).

In section~\ref{sec:stat} we will need the Euler-Lagrange
equations and the Hessian of \refeq{vact}. A
straightforward computation~\cite{AlFrMu} gives (for vanishing
$\delta v(x_1\rightarrow\pm\infty,x_2)$):
\bea
S[v+\delta v]&=&S[v]+S^{(1)}[v,\xi]+S^{(2)}[v,\xi]+\cO(\xi^3),
\nonumber\\
S^{(1)}[v,\xi]&=&-\intx \left(D_{\mu}\dmu v\right)_i\xi^i,
\label{eq:pertact}\\
S^{(2)}[v,\xi]&=&\intx \xi^i\cH_{ij}[v]\xi^j,\ \ \
\cH_{ij}[v]=-\half\left(D_{\mu}D_{\mu}\right)_{ij}+\half R_{kijl}\dmu
v^k\dmu v^l.
\nonumber
\eea
Here $\xi=\delta v+\cO(\delta v^2)$ is defined in such a way that it
transforms covariantly; see~\cite{AlFrMu} for details. Since the
action is a scalar, this guarantees that \cH\ is a tensor, as can be
verified from the explicit form. Furthermore, $D_{\mu}$ is
the covariant derivative and $R$ is the Riemann tensor, both at the
point $v$:
\bea
\left(D_{\mu}\lambda\right)_i&=&\dmu\lambda_i-{\Gamma^j}_{ik}\dmu
v^k\lambda_j,\ \ \ \mbox{(for any vector $\lambda$)} \nonumber\\
\Gamma_{kij}&=&\half\left(\hat{\partial}_j g_{ik}+\hat{\partial}_i
g_{jk}-\hat{\partial}_k g_{ij}\right),\ \ \
(\hat{\partial}_i\equiv\frac{\partial}{\partial v^i})
\\
{R^l}_{ijk}&=&\hat{\partial}_j{\Gamma^l}_{ik}-\hat{\partial}_k
{\Gamma^l}_{ij}
+{\Gamma^m}_{ik}{\Gamma^l}_{mj}-{\Gamma^m}_{ij}{\Gamma^l}_{mk}.
\nonumber
\eea
The Euclidean action, \refeq{vact}, is precisely half the
total volume of \mbox{$v(\tore)\subset S^2$}.
This action also naturally
occurs in string theory (see e.g.~\cite{GrScWi}). Therefore,
extremizing the action (which amounts to putting $D_{\mu}\dmu v=0$,
\refeq{pertact}) gives a geodesic surface (in affine parametrization)
on the space $S^2$.

Finally, it is important to introduce the winding number which
measures the number of times \tore\ is wrapped around $S^2$ by
$v$ (and therefore is invariant under continuous deformations
of $v$):
\be
Q[v]=-\ethpi\intx \varepsilon_{\mu\nu}\vn\cdot\left(\dmu\vn
\times\dnu\vn\right)=\ethpi\intx\varepsilon_{\mu\nu}\dmu\vn\cdot\left(
\vn\times\dnu\vn\right)=\ethpi\intx\varepsilon_{\mu\nu}g_{ij}
\dmu v^i\lambda^j_{\nu},
\label{eq:topch}
\ee
where $\lambda^j_{\nu}$ is the vector in the tangent space of $S^2$
(at the point $v$) corresponding to $\vn\times\dnu\vn$; up to orientation
$\lambda_{\nu}$ is defined by $g_{ij}\lambda^i_{\nu}\lambda^j_{\nu}=
g_{ij}\dnu v^i\dnu v^j,\;g_{ij}\lambda^i_{\nu}\dnu v^j=0$ (no summations
over $\nu$). From this follows the well-known formula~\cite{BePo}
\be
S=\tilde{S}_{\pm}\mp4\pi Q,\ \ \ \tilde{S}_{\pm}[v]=
\fourth\intx g_{ij}\left(\dmu v^i\pm\varepsilon_{\mu\nu}
\lambda^i_{\nu}\right)
\left(\dmu v^j\pm\varepsilon_{\mu\rho}\lambda^j_{\rho}\right).
\label{eq:niceact}
\ee
We define an instanton to be a minimal action configuration in the
sector $Q=+1$. Therefore it has action $4\pi$ and satisfies the instanton
equation
\be
\dmu v^i=\varepsilon_{\mu\nu}\lambda^i_{\nu}.
\label{eq:insteq}
\ee

\section{Sphalerons, vacua and other static solutions}
\label{sec:stat}

In this section we study the potential, or energy functional,
which is the action (\ref{eq:vact}) restricted to space, i.e.\
without time-dependence and without integration over time:
\be
V[v]=\half\int_{{\rm T}^1}\!\!dx_2\,\, g_{ij}(v)\partial_2 v^i
\partial_2 v^j.\ \ \ (v=v(x_2))
\label{eq:pot}
\ee
This is the geodesic action for a curve $v(x_2)$ on $S^2$.
Therefore all static solutions of
the Euler-Lagrange equations, being extrema of \refeq{pot}, are big
circles on $S^2$ (affinely parametrized by $x_2$). Remembering
the requirement $v(x_2+2\pi)=v(x_2)$ we conclude that the most
general static solution reads (after an \soth~rotation)
\be
\vn_k(x_2)=\left(\begin{array}{c}\cos(kx_2) \\
\sin(kx_2) \\ 0\end{array}\right),\ \ \ k\in\zahlen.
\label{eq:statsol}
\ee
One easily computes the energy
\be
V[\vn_k]=\pi k^2.
\label{eq:staten}
\ee
In particular $\vn_0$ is a classical vacuum. Due to \soth~symmetry
the vacuum valley is isomorphic to $S^2$.

Now we turn to the sphalerons. For this we have to determine the Hessian
of the energy functional, \refeq{pot}, at the field
$\vn_k$, \refeq{statsol}. This is easy in spherical coordinates where
\refeq{statsol} reads
$\vartheta_k(x_2)=\frac{\pi}{2},\;\varphi_k(x_2)=kx_2$ and the metric is
given by $(g_{ij})={\rm diag}(1,\sin^2\vartheta)$.
Substituting these formulas in the static version of \refeq{pertact}
we obtain
\be
\cH[\vn_k]=\half\left(\begin{array}{cc}-\partial_2^2-k^2 & 0 \\ 0 &
-\partial_2^2\end{array}\right).
\ee
One immediately sees that
$k=\pm1$ is the only solution with exactly one unstable mode
($\delta\vartheta(x_2)=1$, $\delta\varphi(x_2)=0$).
So\footnote{At this point we do not prove that these saddle points are
true sphalerons in the mini-max sense. This proof
can be constructed easily with the results of the next section.}
 the sphaleron
solutions are given by $\vn_1$
and \soth~transformations thereof. In particular $k\rightarrow-k$ can
be undone by a rotation over $\pi$ around any vector
$\vn_k(x_2)$ ($x_2$ fixed). Thus, sphaleron moduli-space
is isomorphic to $\soth$.

One can verify that
\soth~rotations are responsible for the 3 zero-modes of the
Hessian. Note that the sphaleron is invariant
under an $x_2$-translation in combination with a specific
${\rm SO}(2)\subset\soth$~rotation (in the case of \refeq{statsol} around
the 3-axis). Therefore spatial translations do not give new sphaleron
solutions. Also one can check that the discrete symmetries
$x_2\rightarrow-x_2$ and $\vn\rightarrow-\vn$ leave the sphaleron
moduli-space invariant.

\section{Instanton solutions}
\label{sec:inst}
\subsection{Construction of the solutions}

In stereographic projection~\cite{BePo,RiRo} the instanton equation
(\ref{eq:insteq}) is particularly simple:
\be
\dbz u=0.
\label{eq:analyt}
\ee
Here we have introduced complex coordinates on space-time, $z=x_1+ix_2$,
$\bar{z}=x_1-ix_2$, with derivatives $\dz=\half(\partial_1-i\partial_2)$,
$\dbz=\half(\partial_1+i\partial_2)$. The construction of instantons
reduces to finding
all analytic functions on \tore\ with topological charge $Q=1$.

Substitution of \refeq{npar} in \refeq{topch} gives, for any $u$
satisfying \refeq{analyt},
\be
Q[u]=\fstpi\intxD{\torekl\,}\frac{|\dz u|^2}{(1+|u|^2)^2}=
-\fstpi\intxD{\torekl\,}\dbz\left(\frac{1}{1+|u|^2}\frac{\dz u}{u}\right).
\label{eq:utopch}
\ee
Handling carefully the poles of $\frac{1}{1+|u|^2}\frac{\dz u}{u}$,
i.e.\ the zeros of $u$, one finds
\be
Q[u]=-\hpi\left(\int_{x_1=\infty} - \int_{x_1=-\infty}\right)dx_2\,\,
\frac{1}{1+|u|^2}\frac{\dz u}{u}+\sum_i n_i,
\label{eq:nicetopch}
\ee
where the $x_2$~integrations run from $0$ to $2\pi$ and $i$ runs over
the zeros of $u$, of degree $n_i\in\na$. In order to simplify this
formula, we observe that
both the kinetic and the potential term in the Lagrangian
are semi-positive definite. Hence any finite-action configuration, in
particular an instanton, must
approach a point in the vacuum valley for $x_1\rightarrow\pm\infty$:
$\lim_{x_1\rightarrow\pm\infty}u(z)=u_{\pm}$. In the derivation below
we will assume that
$0<|u_{\pm}|<\infty$, as can always be achieved by an
\soth~rotation\footnote{This is equivalent to changing coordinates
on $S^2$ by shifting the pole used in stereographic
projection.}. Under this
assumption \refeq{nicetopch} reduces to $Q[u]=\sum_i n_i$.

Now we are fully prepared to determine all instanton solutions. The
strategy is first to find a class of solutions and then to prove no
solutions exist outside this class. In order to construct a
solution observe that
\begin{enumerate}
\item Since $e^{z+2\pi i}=e^z$, any $u=h(e^z)$ ($h$ single-valued and
analytic) is a function on \tore\ satisfying the instanton equation.
\item Under the above assumption any instanton can have only one zero
$z_1$ of degree $n_1=1$.
\end{enumerate}
A class of functions satisfying all requirements is
\be
u^{\rm inst}_{a,b,c,d}(z)=-\frac{c+d e^z}{a+b e^z},
\label{eq:inst}
\ee
with certain restrictions on the complex coefficients $a,b,c,d$; note
that $u_{+}=-d/b$, $u_{-}=-c/a$ and $z_1=\ln(-c/d)$ (which is unique on
\tore). So in order to satisfy the assumption $0<|u_{\pm}|<\infty$, we
must require $a,b,c,d\not=0$. Also we must demand $a/b\not=c/d$ as
otherwise the zero of $u^{\rm inst}_{a,b,c,d}$ cancels against its
pole and we have a trivial solution with $Q=0$.

To prove that all instanton solutions are of the form (\ref{eq:inst}) is
easy: suppose $\tilde{u}$ is an instanton. We can still assume
$0<|\tilde{u}_{\pm}|<\infty$, so $\tilde{u}$ can have only
one zero of degree 1, say at $z=\tilde{z}_1$. Now define a function
$f$ by
\be
\tilde{u}(z)=u^{\rm inst}_{a,b,c,d}(z)f(z).
\nonumber
\ee
By choosing $-c/a=\tilde{u}_{-}$, $-d/b=\tilde{u}_{+}$,
$c/d=-e^{\tilde{z}_1}$ and imposing the
instanton requirements $\dbz \tilde{u}=0$, $Q[\tilde{u}]=1$ and
$\tilde{u}(z+2\pi i)=\tilde{u}(z)$, that are already met by $u$, we
see that $f$ has to satisfy
\be
\left\{ \begin{array}{l}\dbz f=0 \\ \lim_{{\rm Re}(z)\rightarrow\pm
\infty}f(z)=1 \\ f(z+2\pi i)=f(z) \\ f(z)\not=0.\end{array}\right.
\label{eq:fisone}
\ee
Thus, $1/f$ is analytic and bounded on \co. By Liouville's theorem this
implies, using the second condition in
\refeq{fisone}, that $f(z)=1$. This completes the proof.

Finally we drop the assumption $0<|u_{\pm}|<\infty$.
Taking into account the boundary terms in
\refeq{nicetopch}, one finds that the only restriction on $(a,b,c,d)$ is
that $u^{\rm inst}_{a,b,c,d}(z)$ is not constant, corresponding to
\be
ad-bc\not=0.
\label{eq:coefreq}
\ee
So the class of instantons, \refeq{inst}, is precisely the set of
conformal mappings of $e^z$.

We end this part by noting that by the same line of argument each
multi-instanton with topological charge $Q$ can be written as
$\prod_{n=1}^Q u^{\rm inst}_{a_n,b_n,c_n,d_n}$. For
(multi-)anti-instantons one substitutes
$z\rightarrow\bar{z}$.

\subsection{Physical interpretation of the moduli-space}

It is clear that the moduli-space has six real dimensions: $(a,b,c,d)\in
\co^4$ with the projective character $u^{\rm
inst}_{ga,gb,gc,gd}=u^{\rm inst}_{a,b,c,d}$ for any $g\in\co\backslash
\{0\}$ (while the set of coefficients not satisfying \refeq{coefreq} has
zero measure). A physical parametrization of this moduli-space is
$(-c/a,-d/b,\ln(-c/d))$, corresponding to the starting point $u_{-}$ in
the
vacuum valley $S^2$, the end point $u_{+}$ and a space-time translation
parameter, which one can show to be in $1-1$~relation with the
instanton position (cf. \refeq{instsol} below). The
disadvantage of this parametrization is that it leaves unclear what kind
of manifold the moduli-space is; the parametrization is singular for
$u_{-}=u_{+}$, since the requirement of \refeq{coefreq} is not met in
this case. Note that this means that even for $T\rightarrow\infty$
periodic boundary conditions do not admit instanton solutions. We
will see below that $u_{-}\rightarrow u_{+}$ corresponds to
$\rho\rightarrow0$, where $\rho$ is the instanton scale parameter.

For a better description of moduli-space it is necessary first to
consider the transformation of instantons under space-time translations
and \soth~rotations. We will see that most
instantons are not invariant under these symmetries which therefore give
rise to five of the six dimensions of moduli-space. The interesting
sixth parameter, not related to a symmetry of the action, will play the
role of a scale parameter, related to the geodesic distance between the
points $u_{\pm}$ in the vacuum valley $S^2$.

{}From \refeq{inst} it is trivial to see that under a translation
$z\rightarrow z+z_0$, $(a,b,c,d)\rightarrow(a,b e^{z_0},c,d e^{z_0})$.
Note that \refeq{coefreq} is therefore invariant under this shift, as
it should be. Any continuous symmetry of the action must be present
in the instanton moduli-space.

The effect of an \soth~rotation is more difficult to derive, because
while it acts linearly on \vn, \vn\ and $u$ are related non-linearly
by \refeq{npar}. Nevertheless, one can show
that $(a,b,c,d)$ again transform linearly. After some effort one sees
that the rotation
\be
\vnx\rightarrow R\vnx,\ \ \ R=e^{\alpha_a L^a},\ \ \ {L^a}_{ij}=
-\varepsilon_{aij},\ \ \ (\alpha_a\in\re)
\ee
induces
\be
\mbox{\scriptsize $\left(\!\!\begin{array}{c}a\\ b\\ c\\ d\end{array}
\!\!\right)$}\rightarrow
\tilde{R}\mbox{\scriptsize $\left(\!\!\begin{array}{c}a\\ b\\ c\\ d
\end{array}\!\!\right)$},\ \ \
\tilde{R}=e^{\alpha_a \tilde{L}^a},\ \ \ \tilde{L}^a=-\frac{i}{2}\sigma^a
\otimes{\bf 1},\ \ \ {\rm (i.e.\ } \tilde{L}^1= -\frac{i}{2}
\mbox{\scriptsize $\left(\!\!
\begin{array}{cccc}0&0&1&0\\ 0&0&0&1\\ 1&0&0&0\\
0&1&0&0\end{array}\!\!\right)$} {\rm\ etc.)}
\label{eq:rot}
\ee
where $\sigma^a$ are the Pauli-matrices. Since $\pm(a,b,c,d)$ are
identified, this is a representation of \soth. Notice that only
$(a,c)$ and $(b,d)$ mix. Hence both $|a|^2+|c|^2$ and
$|b|^2+|d|^2$ are rotationally invariant, as is \refeq{coefreq}.

Using the projective character of moduli-space, and an \soth~rotation
$\tilde{R}$, it is always possible to bring
$u^{\rm inst}_{a,b,c,d}$ ($ad\not= bc$) to the form
$u^{\rm inst}_{-1,0,\tilde{c},\tilde{d}}$ with $\tilde{c}\in\re$,
$\tilde{c}\geq0$, $\tilde{d}\in\co\backslash\{0\}$. If $\tilde{c}>0$,
this fixes $\tilde{R}$ completely. If
$\tilde{c}=0$, then $\tilde{R}$ is only unique up to a factor
$e^{\alpha_3\tilde{L}^3}$. This can be fixed by requiring
$\tilde{d}=|\tilde{d}|$. Therefore, we can parametrize
$u^{\rm inst}_{a,b,c,d}$ uniquely by $\tilde{R}$, $|\tilde{d}|$ and
$\tilde{c}e^{i\phi}$ ($\tilde{c}\geq0$). Here $\phi={\rm
Arg}(\tilde{d})$ ($\phi\in[0,2\pi)$) if $\tilde{c}>0$ and $\phi$ is
undetermined if $\tilde{c}=0$. We conclude that $(\tilde{c},\phi)$ can
be viewed as polar coordinates on $\re^2$. Furthermore, $|\tilde{d}|
>0$ due to \refeq{coefreq}, so $\ln|\tilde{d}|\in\re$. Thus the instanton
moduli space is isomorphic to $\soth\times\re^2\times\re$. The discrete
symmetry transformations $\vn\rightarrow-\vn$
(in stereographic coordinates $u\rightarrow-1/\bar{u}$),
$x_2\rightarrow-x_2$ or $x_1\rightarrow-x_1$ make an instanton solution
anti-analytic, and therefore are transformations from instanton
moduli-space into anti-instanton moduli-space. This should be compared
to the sphaleron moduli space which is invariant under such discrete
transformations.

Note that $\tilde{d} e^z=e^{z+\ln|\tilde{d}|+i\phi}$. So after an
\soth~rotation and a translation in space-time \tore, any instanton
solution can be brought to the form
\be
u^{\rm inst}_c(z)\equiv c+e^{z+\ln(\sqrt{1+c^2})+i\pi}.\ \ \ (c\geq0)
\label{eq:instsol}
\ee
The factor $e^{\ln(\sqrt{1+c^2})+i\pi}=-\sqrt{1+c^2}$ centers the
instanton around $z=0$ (see below). From the above equation it
follows that $\lim_{x_1\rightarrow\infty}|u^{\rm
inst}_c(z)|=\infty$ and $\lim_{x_1\rightarrow-\infty}u^{\rm
inst}_c(z)=c$, hence
(using \refeq{npar}) this instanton `tunnels' from
$(\sin\vartheta_{-},0, \cos\vartheta_{-})$ to $(0,0,1)$, with
$\cot\half\vartheta_{-}=c$. Note that
$\theta_{-}$ is the geodesic distance between these vacua.

\begin{figure}[t]
\epsfxsize=\textwidth
\epsffile{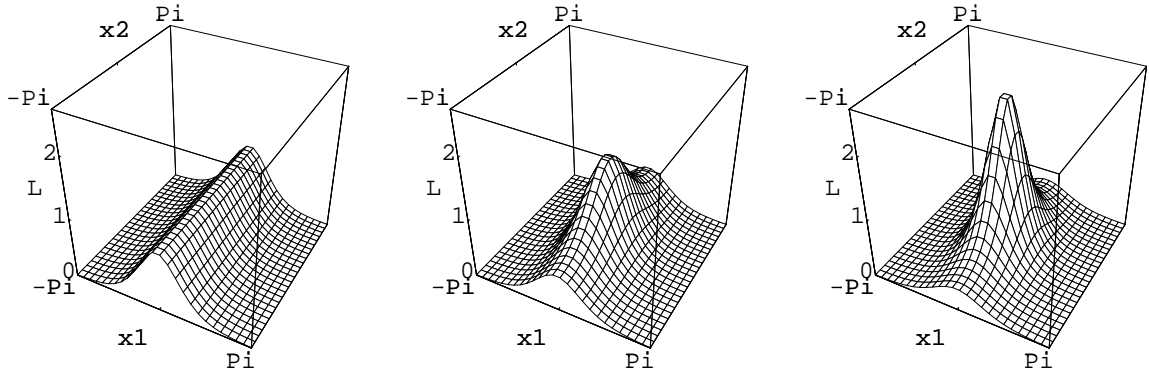}
\caption{The Lagrangian densities of three instantons,
\protect\refeq{instsol},
with from left to right $c=0,\:0.25,\:0.5$.}
\label{fig:lagr}
\end{figure}

Let us determine the instanton size $\rho$ as function of
$c=2{\rm arccot}\vartheta_{-}$. Substituting \refeq{instsol} into the
Lagrangian density, which for an instanton in stereographic coordinates
is $4\pi$ times the integrand in \refeq{utopch} (see
eqs.(\ref{eq:niceact},\ref{eq:insteq})), one obtains
\be
\cL[u^{\rm inst}_c](x_1,x_2)=4\frac{|\dz u^{\rm inst}_c|^2}{(1+|u^{\rm
inst}_c|^2)^2}=\frac{1}{(\sqrt{1+c^2}\cosh x_1-c\cos x_2)^2}.
\label{eq:instlagr}
\ee
This function is plotted in fig.\ref{fig:lagr} for different values of
$c$. From the formula it is clear that $u^{\rm inst}_c$ is
centered at $z=0$. Now consider the potential along the instanton path,
\be
V[u^{\rm inst}_c](x_1)=\half\int_0^{2\pi}\!\!
dx_2\,\,\cL[u^{\rm inst}_c](x_1,x_2).
\ee
The factor $\half$ comes from the fact that the kinetic energy,
$\half\int_{{\rm T}^1}\!dx_2\, g_{ij}(v)\partial_1v^i\partial_1v^j$,
is equal to the potential energy, \refeq{pot} (this `self-duality'
follows from \refeq{insteq}). We see that the potential is maximal at
$x_1=0$ where it satisfies
\be
V^{\max}_c\equiv V[u^{\rm inst}_c](0)=\pi\sqrt{1+c^2}.
\label{eq:insten}
\ee
Since all instantons have equal action, it is natural to define the
instanton size
\be
\rho(c)\equiv\frac{\pi}{V^{\max}_c}=\frac{1}{\sqrt{1+c^2}}.
\label{eq:instrho}
\ee
Note however that for small $c$ there are two different scales; the
shape of $\cL[u^{\rm inst}_c](x_1,x_2)$ is anisotropic (see
fig.\ref{fig:lagr}). Only for $c\gg1$ and $x_1^2+x_2^2\ll1$ the
boundary effects disappear and $\cL[u^{\rm inst}_c](x_1,x_2)\approx
\frac{4c^2}{(1+c^2(x_1^2+x_2^2))^2}$ becomes rotationally invariant.

The relationship between instantons and sphalerons is now also clear.
{}From eqs.(\ref{eq:insten},\ref{eq:staten}) we see that only
$u^{\rm inst}_{c=0}$ can go through a sphaleron at the time of maximal
$V[u^{\rm inst}_c]$ (i.e.\ $x_1=0$). Indeed this does happen, since
\refeq{instsol} gives $u^{\rm inst}_{c=0}(x_1=0,x_2)=-e^{ix_2}$,
which up to a rotation is just the sphaleron solution $\vn_1$,
\refeq{statsol},
in stereographic coordinates. As the instanton is a self-dual solution
of the equations of motion, it has to follow streamlines of the energy
functional. This explains why
$\partial_1 u^{\rm inst}_{c=0}|_{x_1=0}=-\exp(ix_2)$
corresponds to the unstable mode of the sphaleron solution.

Finally note that
$u^{\rm inst}_{c=0}$, unlike $u^{\rm inst}_{c>0}$, is a point in
instanton moduli-space that is symmetric under a joint
${\rm SO}(2)$~rotation $e^{\alpha_{3}\tilde{L}^3}$ and a spatial
translation $x_2\rightarrow x_2-\alpha_3$. The sphaleron has of
course the same symmetry, as mentioned in section~\ref{sec:stat}.
Also a natural correspondence between sphaleron moduli-space \soth\
and the subspace of widest instantons emerges. From the paragraph
above \refeq{instsol} it follows that the latter is isomorphic to
$\soth\times\re$, $\re$ corresponding to time translations.

\section{Conclusions}
\label{sec:concl}

We have proven that the \nlsig\ on a space-time \tore\ admits instantons.
The moduli-space is 6-dimensional ($\soth\times\re^2\times\re$):
3 parameters for \soth, 2 for scaling and spatial translations, 1 for
time translations. It is possible to `tunnel' between any different
points $\vn_{\pm}$ in the vacuum valley ($S^2$),
but this gives no independent parameters. Three parameters describing
$\vn_{\pm}$ can be removed by an \soth~rotation, while
the fourth (the geodesic distance between the points) depends
uniquely on the scale parameter $\rho$. Instantons with maximum scale, as
set by the extent of spatial ${\rm T}^1$, satisfy $\vn_{+}=-\vn_{-}$. These,
and only these, instantons go through sphalerons. None of the exotic
possibilities sketched in the introduction take place in this model.
On the other hand,
$\rho\rightarrow0$ corresponds to $\vn_{+}\rightarrow\vn_{-}$. Exact
equality cannot be reached. We think this peculiar size-dependence
is important for improving on instanton gas calculations as
in~\cite{BeLu}. By doing a proper convolution with the vacuum wave
function at $t\rightarrow\pm\infty$ it might be possible to remove the
well-known UV divergence for $\rho\rightarrow0$, which was also
encountered in recent numerical studies~\cite{MiSp}.

Our results do not admit a straightforward generalization to
\sutw~gauge theory on a space-time \threetore. In that model
the vacuum valley is isomorphic to \thrto~\cite{BaKo},
which can be parametrized
by three Polyakov lines $P_i$. Instantons again must have endpoints,
$P_i^{\pm}$, in the vacuum valley. For the special case
$P_i^{+}=-P_i^{-}$ it has already been known for some time
that an 8-dimensional moduli-space exists. This anti-periodic
situation corresponds to time-like twist~\cite{Ho}, which has been
analyzed~\cite{BrMaTo} on any space-time $\intT\times\thrto$. For
$T\rightarrow\infty$ it is very likely~\cite{PeGoSnBa} that the
moduli-space includes a scale parameter. This is not the case
for the \nlsig\ on a space-time \tore. We have just proven that
anti-periodic boundary conditions, $\vn_{+}=-\vn_{-}$, fix the
instanton size\footnote{We also note that no instantons exist
with anti-periodic boundary conditions in finite time $T$,
\mbox{$\vn(x_1+T,x_2)=-\vn(x_1,x_2)$}
(while the winding number $Q$ is still a well-defined integer object).
The reason is simple: in stereographic coordinates anti-pbc read
$u(z+T)=-1/\bar{u}(z)$, which is incompatible with the instanton
equation $\dbz u=0$.}. It would be nice to understand the cause of
such different behavior
between two models that are so similar in other respects. This might
be a starting point for finding new instanton parameters in \sutw~gauge
theory on \threetore, by relaxing the condition $P_i^{+}=-P_i^{-}$.

\section{Acknowledgments}

I am grateful to Pierre van Baal for fruitful discussions and for
carefully reading this manuscript. Also I thank Pieter Rijken for
expanding my knowledge of analytic functions, and Margarita
Garc\'{\i}a P\'{e}rez for discussions on various sphaleron-related
topics.


\begin{thebibliography}{99}
\bibitem{LuKoBa}
M. L\"{u}scher, Nucl.Phys. B219 (1983) 233;\\
J. Koller and P. van Baal, Nucl. Phys. B302 (1988) 1.
\bibitem{BaDa}
P. van Baal and N.D. Hari Dass, Nucl.Phys. B385 (1992) 185.
\bibitem{Ta}
C.H. Taubes, Comm. Math. Phys. 86 (1982) 257, 299;
C.H. Taubes, in: Progress in gauge field theory, eds. G. 't Hooft
et. al. (Plenum Press, New York, 1984) p.~563.
\bibitem{KlMa}
F.R. Klinkhamer and N. Manton, Phys.Rev. D30 (1984) 2212.
\bibitem{BePo}
A.A. Belavin and A.M. Polyakov, J.ETP-Lett. 22 (1975) 245.
\bibitem{BePoScTy}
A.A. Belavin, A.M. Polyakov, A.S. Schwartz and Yu.S. Tyupkin,
Phys.Lett. B59 (1975) 85;
G. 't Hooft, Phys. Rev. D14 (1976) 3432.
\bibitem{PeGoSnBa}
M. Garc\'{\i}a P\'{e}rez, A. Gonz\'{a}lez-Arroyo, J. Snippe and P. van
Baal, Nucl.Phys. B413 (1994) 535.
\bibitem{GaBa}
M. Garc\'{\i}a P\'{e}rez and P. van Baal, Sphalerons and other Saddles
from Cooling, Leiden preprint INLO-PUB-2/94 (1994).
\bibitem{BrBa}
P.J. Braam and P. van Baal, Comm. Math. Phys. 122 (1989) 267.
\bibitem{RiRo}
J.-L. Richard and A. Rouet, Nucl.Phys. B211 (1983) 447.
\bibitem{MiSp}
C. Michael and P.S. Spencer, Instanton Size Distribution in O(3),
Liverpool preprint LTH331 (1994).
\bibitem{AlFrMu}
L. Alvarez-Gaum\'{e}, D.Z. Freedman and S. Mukhi, Ann. Phys. (N.Y.)
134 (1981) 85.
\bibitem{GrScWi}
M. Green, J.H. Schwartz and E. Witten, Superstring theory, Vol.1
(Cambridge University Press, 1987).
\bibitem{BeLu}
B. Berg and M. L\"{u}scher, Comm. Math. Phys. 69 (1979) 57.
\bibitem{BaKo}
P. van Baal and J. Koller, Ann. Phys. (N.Y.) 174 (1987) 299.
\bibitem{Ho}
G. 't Hooft, Nucl. Phys. B153 (1979) 141.
\bibitem{BrMaTo}
P. Braam, A. Maciocia and A. Todorov, Invent. Math 108 (1992) 419.
\end{thebibliography}
\end{document}